\documentclass[aps,pra,twocolumn,groupedaddress,showpacs,amsmath]{revtex4}

\usepackage{graphics}

\newcommand*{\br}{\mathbf{r}}
\newcommand*{\bp}{\mathbf{p}}
\newcommand*{\bk}{\mathbf{k}}
\newcommand*{\bq}{\mathbf{q}}

\newcommand*{\cC}{{\cal C}}

\begin{document}

\title{Particle correlations in a fermi superfluid}

\author{A. Lamacraft}
\affiliation{Rudolf Peierls Centre for Theoretical Physics,1 Keble Road, Oxford OX1 3NP, UK
and All Souls College, Oxford.}
\date{\today}
\email{austen.lamacraft@all-souls.ox.ac.uk}

\begin{abstract}

We discuss correlations between particles of different momentum in a superfluid fermi gas, accessible through noise measurements of absorption images of the expanded gas. We include two elements missing from the simplest treatment, based on the BCS wavefunction: the explicit use of a conserving approximation satisfying particle number conservation, and the inclusion of the contribution from Cooper pairs at finite momentum. We expect the latter to be a significant issue in the strongly correlated state emerging in the BCS-BEC crossover.

\end{abstract}


\maketitle

The recent observation of resonance pairing in gases of fermionic atoms in the BCS-BEC crossover regime~\cite{regal2004} has been followed by an explosion in experimental and theoretical activity aimed at understanding this canonical problem of many-body physics. The measurement of the paring gap~\cite{chin2004} and most recently the observation of superfluid flow~\cite{zwierlein2005} have dramatically confirmed the presence of superfluid condensates in these systems.

One experiment that is yet lacking is a direct measurement of \emph{order}, that is, a demonstration of condensation of pairs into the a zero momentum state. In the case of Bose-Einstein condensation of bosonic atoms (or molecules), the distribution $n(\bk)$ of the number of particles in each momentum state is readily available by time-of-flight imaging. The signature of BEC is a sharp peak in this distribution at zero momentum. By contrast, $n(\bk)$ in a fermionic superfluid is a smooth function that resembles the corresponding distribution in a normal fermi gas at finite temperature -- see the recent experiment~\cite{regal2005}. Thus such a  one-particle measure is of little use.

A ingenious resolution to this problem was the suggestion in Ref.~\cite{altman2004} to measure \emph{noise correlations} in the time-of-flight images. The density-density correlation function of the gas following expansion can be interpreted as the (column-integrated) correlator of $n(\bk)$ at two different momenta in a gas prior to release from the trap. Then the condensation of pairs at zero momentum corresponds to a peak in the correlator
\begin{eqnarray} \label{cdef}
\cC_{\uparrow\downarrow}(\bk_1,\bk_2)&\equiv&\langle n_{\uparrow}(\bk_1)n_{\downarrow}(\bk_2)\rangle-\langle n_{\uparrow}(\bk_1)\rangle\langle n_{\downarrow}(\bk_2)\rangle
\end{eqnarray}
at opposite momenta $\bk_1=-\bk_2$. We denote the two atomic species between which pairing occurs as $s=\uparrow,\downarrow$, and assume species-selective imaging. In Ref.~\cite{altman2004} this correlation function was evaluated  (at zero temperature) using the usual BCS ground state ansatz
\begin{equation}\label{BCS}
|\mathrm{BCS}\rangle=\prod_\bk \left(u_\bk+v_\bk a^{\dagger}_{\uparrow\bk}a^{\dagger}_{\downarrow-\bk}\right)|0\rangle
\end{equation}
to give
\begin{equation}\label{C_naive}
\cC^{\mathrm{BCS}}_{\uparrow\downarrow}(\bk_1,\bk_2)=\delta_{\bk_1,-\bk_2}u_{\bk_1}^2v_{\bk_1}^2=\delta_{\bk_1,-\bk_2}\frac{|\Delta|^2}{4E_{\bk_1}^2},
\end{equation}
where $E_p=\sqrt{\xi_p^2+|\Delta|^2}$, and $\xi_\bp=\bp^2/2m-\mu$. The first experiments demonstrating the possibility of such measurements have now been performed~\cite{greiner2005,folling2005}. Since a measurement of $\cC_{\uparrow\downarrow}(\bk_1,\bk_2)$ in the BCS-BEC crossover may be attempted soon, and given the utility of such a measurement in constraining many-body theories of the strongly interacting gas, our purpose in this paper is to develop the theory of noise correlations in the following two directions:

\emph{1. Conserving approximation}. We note that Eq.~(\ref{C_naive}), together with the correlator $\cC_{\uparrow\uparrow}(\bk_1,\bk_2)$, which has the same form but with a delta-function $\delta_{\bk_1,\bk_2}$, can be used to calculate the density structure factor $S(\bq)=\langle \rho_\bq\rho_{-\bq}\rangle$ of the system at zero momentum. The result (\ref{C_naive}) gives
\begin{eqnarray} \label{struct_fact_naive}
S(0)&=&2\sum_{\bk_1,\bk_2}\cC_{\uparrow\downarrow}^{\mathrm{BCS}}(\bk_1,\bk_2)+\cC_{\uparrow\uparrow}^{\mathrm{BCS}}(\bk_1,\bk_2)\nonumber\\&=&\nu\int d\xi\frac{|\Delta|^2}{E_{\bk_1}^2}=\pi\nu |\Delta|, 
\end{eqnarray}
(in the BCS limit, where $\nu$ is the density of states per atomic species at the fermi energy) However, it is well known~\cite{price1954} (see~\cite{buchler2004} for a recent discussion in the context of fermi gases) that the structure factor of a quantum fluid with finite compressibility \emph{vanishes} at zero momentum at zero temperature. The incorrect result (\ref{struct_fact_naive}) first appeared in~\cite{anderson1958}, where it was used to conclude that the low energy modes of a neutral superconductor would have dispersion $\omega\sim q^2$, rather than the linearly dispersing Bogoliubov-Anderson mode. Since (\ref{struct_fact_naive}) arises entirely from the correlations found in Ref.~\cite{altman2004}, one may worry that the result (\ref{C_naive}) will be changed in some essential way.

Note that this problem is unrelated to the non-conserving form of Eq.~(\ref{BCS}). It is easy to see that the projected form
\[|\mathrm{BCS}\rangle_N=\left(\sum_\bk \frac{v_\bk}{u_\bk} a^{\dagger}_{\uparrow\bk}a^{\dagger}_{\downarrow-\bk}\right)^{N/2}|0\rangle, \]
lacks the long-ranged correlations in position space required to make the structure factor vanish~\footnote{We should point out that the finiteness of $S(\bq)$ at $\bq\to 0$ is not inconsistent with the value exactly at $\bq=0$ being zero, as it will be for a number-conserving wavefunction. The requirement $S(\bq\to 0)\to 0$ is a stronger constraint, not met in general by such wavefunctions.}.

The resolution to this problem, as described in the book~\cite{schreiffer_book}, is to use a \emph{conserving approximation}. In the language of diagrams, we must include vertex corrections required by particle number conservation. Our first goal in this paper is to explain the effect these have on the correlator $\cC_{\uparrow\downarrow}(\bk_1,\bk_2)$. The good news is that the delta-function structure of (\ref{C_naive}) remains intact, but the correlations function acquires a smooth background 
\begin{equation}\label{C_cons}
\cC_{\uparrow\downarrow}(\bk_1,\bk_2)=\cC^{\mathrm{BCS}}_{\uparrow\downarrow}(\bk_1,\bk_2)+\tilde\cC(\bk_1,\bk_2),
\end{equation}
where $\tilde\cC(\bk_1,\bk_2)$ depends only on the magnitude of the two momenta, and is smooth on the scale of the gap. This term cancels the first one when we compute the structure factor as in Eq.~\ref{struct_fact_naive}. It is natural to interpret $\cC(\bk_1,\bk_2)$ as arising from correlations between different pairs so that, as in a single-species BEC, the vanishing of the structure factor at zero wavenumber is due to pairwise correlations between bosons (i.e. the Bogoliubov approximation). Such an interpretation implies that a measurement of \emph{four}-particle correlations would reveal a peak at $\bk_1+\bk_2+\bk_3+\bk_4=0$

\begin{figure} 
\begin{center}
\setlength{\unitlength}{1.5in}
\begin{picture}(1, 0.8)(0,0)
  \put(-0.3,0.9){\rotatebox{-90}{\resizebox{1\unitlength}{!}{\includegraphics{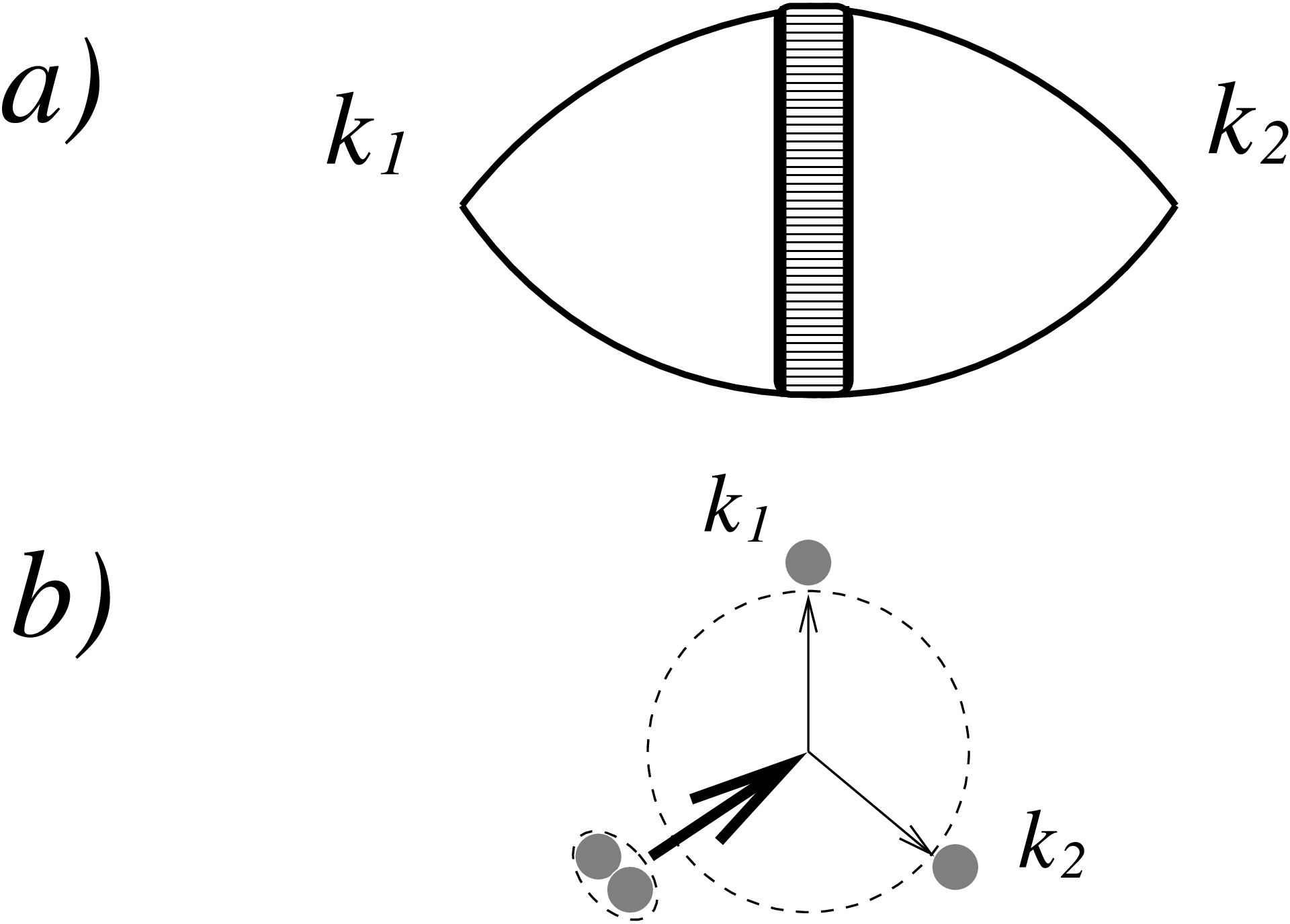}}}}
  \end{picture}
\end{center}
\caption{a) Diagram describing contribution from finite momentum Cooper pairs. b) Schematic of the disintegration process
\label{fig:nq}}
\end{figure}

\emph{2. Cooper pairs at non-zero momentum}. Another well-known consequence of including correlations between bosons is that the distribution function $n(\bq)$ of the number of bosons with given momentum acquires non-trivial structure (leading to condensate depletion).
Thus the true correlation function $\cC_{\uparrow\downarrow}(\bk_1,\bk_2)$ will likely have a strong angular dependence outside of the main peak at $\bk_1=-\bk_2$ due to the presence of such pairs, particularly in the crossover region where we expect deviations from the BCS picture to be most extreme. In fact, part of this dependence turns out to be \emph{universal} and has the form
\begin{eqnarray}\label{nq_result}
N_p\cC_{\uparrow\downarrow}(\bk_1\to-\bk_2)\to\frac{\Delta^2}{4E_{\bk_1}^2}\left[N_p\delta_{\bk_1,-\bk_2}+\frac{(2m)c_s}{2|\bk_1+\bk_2|}\right],\nonumber\\
|\bk_1+\bk_2|\ll \xi^{-1}
\end{eqnarray}
where $N_p=k_F^3/(6\pi^2)\times\mathrm{Vol}$, with $k_F$ the fermi wavevector, is the number of pairs i.e the number of atoms of each species, $c_s$ is the sound velocity, and $\xi$ is the Ginzburg-Landau coherence length that interpolates between the familiar coherence length $k_F/m\Delta$ of the BCS superfluid and the healing length $1/4mc_s$ of the BEC~\cite{engelbrecht1997}. The meaning of this contribution is very simple (see Fig.~\ref{fig:nq}b). Outside of the brackets we have the square of the usual Cooper pair wavefunction $\phi(\bk)=\Delta/E_k$, describing the internal structure of the pairs. Inside the bracket we have a new contribution, which should be interpreted as the the momentum distribution of uncondensed pairs $n_p(\bq)=(2m)c_s/2q$. This form is precisely that expected in simple BEC~\cite{gavoret1964,chester1966}, and can be deduced very generally from the uncertainty principle~\cite{pitaevskii1991}.

The contribution of pairs of finite momentum is already one of the issues in the experiment~\cite{greiner2005}, which measured the correlations of dissociated molecules. The result (\ref{nq_result}) shows how superfluid order manifests itself in this contribution. Clearly this term is small in the extreme BEC limit as $c_s\to 0$. In the BCS limit $c_s$ goes to the finite value $v_F/\sqrt{3}$, where $v_F$ is the Fermi velocity, but the region of validity of (\ref{nq_result}) becomes very small as $\xi\to\infty$. Since $\xi$ attains a minimum value of order the fermi wavelength in the crossover this is a promising place to look for the contribution of finite momentum pairs. It would be very interesting to experimentally confirm these strong deviations from the prediction of the simple BCS wavefunction in the crossover state. 

We now turn to the derivation of these results. Let us first repeat the calculation of Ref.~\onlinecite{altman2004} in the Green's function language to establish our formalism. In terms of fermion creation and annihilation operators the number of particles of each species with momentum $\bk$ is $n_s(\bk)=\psi^{\dagger}_s(\bk)\psi^{\vphantom{\dagger}}_s(\bk)$. To calculate the correlation function (\ref{cdef}) we introduce the Gor'kov Green's functions (real time, zero temperature)
\[\hat G(\bk_1,t_1;\bk_2,t_2)=-i\langle T  \Psi^{\vphantom{\dagger}}(\bk_1,t_1)\otimes \Psi^{\dagger}(\bk_2,t_2)\rangle\]
in terms of the Nambu spinor
\[\Psi(\bk)=\begin{pmatrix}
\psi^{\vphantom{\dagger}}_{\uparrow}(\bk) \cr
\psi^{\dagger}_{\downarrow}(-\bk)
\end{pmatrix} \]
%

\begin{figure} 
\begin{center}
\setlength{\unitlength}{1.5in}
\begin{picture}(1, 1)(0,0)
  \put(-0.3,0.9){\rotatebox{-90}{\resizebox{1\unitlength}{!}{\includegraphics{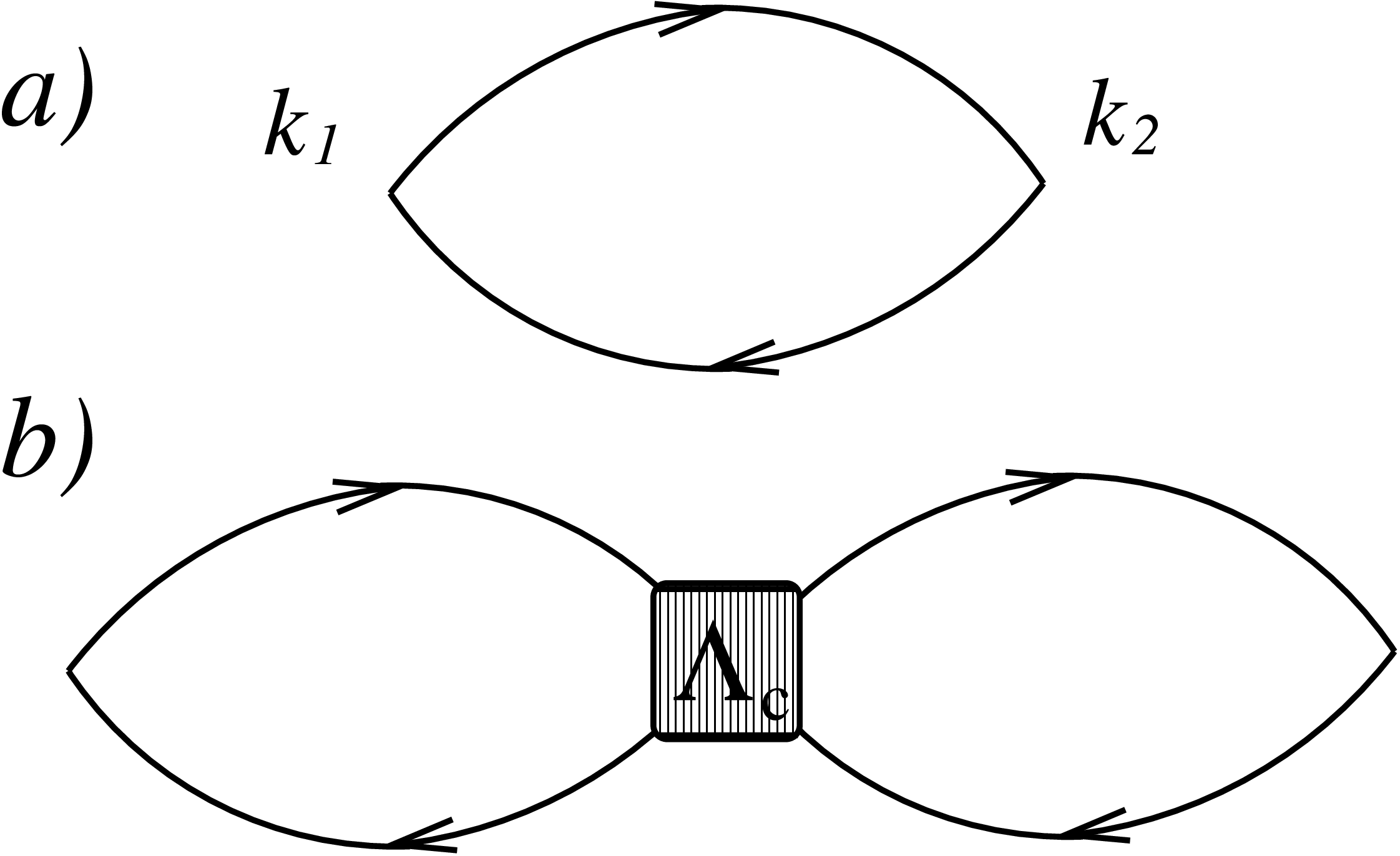}}}}
  \end{picture}
\end{center}
\caption{Diagrams corresponding to a) the naive approximation Eq.~(\ref{C_naive}) and b) the vertex correction required for a conserving approximation.
\label{fig:bubble}}
\end{figure}

Evaluating the correlation function (\ref{cdef}) using Wick's theorem gives
\[\cC_{\uparrow\downarrow}(\bk_1,\bk_2)=-G_{12}(\bk_1,0;-\bk_2,0)G_{21}(-\bk_2,0;\bk_1,0)\]
The diagram corresponding to this contribution is just the bubble (see Fig.~\ref{fig:bubble}). The explicit form of the Green's function is $\hat G(\bk_1,\varepsilon_1;\bk_2,\varepsilon_2)=(2\pi)\delta(\varepsilon_1-\varepsilon_2)\delta_{k_1\bk_2}\hat G_0(\bk_1,\varepsilon_1)$, with
\begin{equation} \label{gf_explicit}
\hat G_0(\bp,\varepsilon)=\frac{1}{\varepsilon^2-E_\bp^2+i\delta}\begin{pmatrix} 
\varepsilon+\xi_\bp && -\Delta\cr
-\Delta^* && \varepsilon-\xi_\bp.
\end{pmatrix}
\end{equation}
Straightforward calculation yields the result (\ref{C_naive}) of Ref.~\onlinecite{altman2004}. The delta-function peak at $\bk_1=-\bk_2$ is the signature of condensation of Cooper pairs with opposing momenta. We could also study the correlations of the same species:
\[\cC_{\uparrow\uparrow}=G_{11}(\bk_1,+;\bk_2,0)G_{11}(\bk_2,0;\bk_1,+)=\delta_{\bk_1,\bk_2}\frac{|\Delta|^2}{4E_{\bk_1}^2}\]
(we have to pay attention to the point splitting in the time direction to get the right answer). The origin of this term is simply the binomial variance $n(\bk)\left(1-n(\bk)\right)$ arising from having $0<n(\bk)<1$ due to the pairing,
%
%
so this part resembles the correlations at finite temperature in a normal fermi gas.

As explained earlier, the result (\ref{C_naive}) cannot be all there is, as it gives the wrong value of $S(0)$. In general, the approximation for a two particle quantity such as $\cC_{\uparrow\downarrow}(\bk_1,\bk_2)$ must be consistent with that used for the self-energy. The self-energy approximation in question is just the usual BCS self-consistency equation. We now describe the resulting approximation for $\cC_{\uparrow\downarrow}(\bk_1,\bk_2)$. Digrammatically, vertex corrections (see Fig.~\ref{fig:bubble}) are required to ensure particle number conservation, i.e. that the vertex function satisfies the generalized Ward identity~\cite{schreiffer_book}. 

Consider the case of the model interaction
\[H_{\mathrm{int}}=\lambda\int d\br \,\psi^{\dagger}_{\uparrow}\psi^{\dagger}_{\downarrow}\psi^{\vphantom{\dagger}}_{\downarrow}\psi^{\vphantom{\dagger}}_{\uparrow}.\]
We can represent this operator in Nambu form in the following way
\begin{eqnarray} \label{int_nambu}
\psi^{\dagger}_{\uparrow}\psi^{\dagger}_{\downarrow}\psi^{\vphantom{\dagger}}_{\downarrow}\psi^{\vphantom{\dagger}}_{\uparrow}&=&\Psi^{\dagger}_1\Psi^{\vphantom{\dagger}}_2\Psi^{\dagger}_2\Psi^{\vphantom{\dagger}}_1=\frac{1}{2}\left(\Psi^{\dagger}\tau_3\Psi\right)\left(\Psi^{\dagger}\tau_3\Psi\right)\nonumber\\&=&\frac{1}{2}\left(\Psi\otimes\Psi^{\dagger}|\tau_3\otimes\tau_3|\Psi^{\dagger}\otimes\Psi\right).
\end{eqnarray}
(some delta-function terms arising from anticommutation drop out). The final form is for later convenience. The simplest approximation for the self-energy corresponding to this interaction is
\begin{equation} \label{se}
\hat\Sigma\left(\bp,\varepsilon\right)=i\lambda\int\frac{d\bp'^3d\varepsilon'}{\left(2\pi\right)^4}\tau_3\hat G\left(\bp',\varepsilon'\right)\tau_3-\tau_3\,\mathrm{tr}\left[\tau_3\hat G\left(\bp',\varepsilon'\right)\right],
\end{equation}
Note that this includes the Hartree and Fock channels as well as the Cooper channel. The Hartree part is the second term of (\ref{se}); the Fock part is included in the first part. The Hartree-Fock potential appears as the coefficient of $\tau_3$ in (\ref{se}), the Cooper part is off-diagonal, and reproduces the usual self-consistent equation. In the following calculation, we will work in the weakly coupled (BCS) limit. For now we drop the Hartree part (second term), and will explicitly demonstrate shortly why the Cooper part is the only one that needs to be accounted for in the weakly coupled problem. 

This first term of Eq.~(\ref{se}) implies that the vertex function should be taken to be the geometric series of bubbles $\hat\Phi(q,\omega)$
\begin{equation}\label{vertex}
\hat\Lambda(\bq,\omega)=\lambda\tau_3\otimes\tau_3\left(1-\lambda\hat\Phi(\bq,\omega)\right)^{-1}
\end{equation}
where
\begin{eqnarray*}
\hat\Phi(\bq,\omega)=-i\int\frac{d^3\bp}{\left(2\pi\right)^3}\frac{d\varepsilon}{2\pi}\hat G_{\bp+\bq/2,\varepsilon+\omega/2}\otimes\hat G^{\mathrm{T}}_{\bp-\bq/2,\varepsilon-\omega/2}
\end{eqnarray*}
The matrix structure of (\ref{vertex}) is inherited from (\ref{int_nambu}). $\hat\Lambda(q,\omega)$ is a $4\times 4$ matrix. We denote the restriction of $\Lambda_{aa'bb'}$ to the subspace $a\neq b'$, $a'\neq b$ as $\hat\Lambda_c$, corresponding to the Cooper channel. The restriction  $\hat\Phi_c(q,\omega)$ is afflicted with the usual log-divergence. To handle this, we use the gap equation
\begin{equation*}
\frac{2}{\lambda}=-\int \frac{d^3\bp}{\left(2\pi\right)^3}\frac{1}{\sqrt{\Delta^2+\xi_\bp^2}}=-\nu\int d\xi \frac{1}{\sqrt{\Delta^2+\xi^2}}
\end{equation*}
($\nu$ is the fermi surface density of states per spin degree of freedom) to write 
\begin{eqnarray*}
\left[\frac{1}{\lambda}-\hat\Phi(0,\omega)\right]_{c}&=&
\nu\int d\xi \frac{1}{\sqrt{\Delta^2+\xi^2}\left(4\left(\Delta^2+\xi^2\right)-\omega^2\right)}\nonumber\\
&&\qquad\begin{pmatrix}
 -\Delta^2+\omega^2/2 & -\Delta^2 \cr
  -\Delta^2 &  -\Delta^2+\omega^2/2  \cr
\end{pmatrix}\\\nonumber
\end{eqnarray*}
Outside of the Cooper subspace, $\hat\Phi(0,\omega)$ is completely convergent, so that when we taken the limit $\lambda\to 0$ as the cut-off goes to infinity (limit of a delta-function potential), we can treat $\left[1/\lambda-\hat\Phi\right]^{-1}$ as zero outside this subspace. 

The final form of the vertex function is
\begin{eqnarray*}
\hat\Lambda_c(0,\omega)&=&\begin{pmatrix}
-\Delta^2+\omega^2/2 & \Delta^2 \cr
  \Delta^2 &  -\Delta^2+\omega^2/2  \cr
\end{pmatrix}\nonumber\\
&&\qquad\frac{2}{\nu|\omega|\sqrt{4\Delta^2-\omega^2}}\left[\cos^{-1}\left(\sqrt{1-\omega^2/4\Delta^2}\right)\right]^{-1}\nonumber\\
&\equiv&\begin{pmatrix}
\Lambda_1(0,\omega) & \Lambda_2(0,\omega) \cr
\Lambda_2(0,\omega) & \Lambda_1(0,\omega)
\end{pmatrix}
\end{eqnarray*}
Note that the vertex matrix has one eigenvalue $\Lambda_1(0,\omega)-\Lambda_2(0,\omega)$ that goes like $1/\omega^2$ as $\omega\to 0$. This is due to the phase mode. The other corresponds to the amplitude mode.

It is now straightforward to find the vertex correction to $\cC_{\uparrow\downarrow}(\bk_1,\bk_2)$. We have to compute 
\begin{widetext}
\begin{eqnarray} \label{first_order}
&&\frac{i}{2}\sum_{k,k'}\int dt_1 dt_2 \Big\langle \Psi^{\dagger}_1(\bk_1,0)\Psi^{\vphantom{\dagger}}_1(\bk_1,0)\left(\Psi^{\dagger}(\bk,t_1)\otimes\Psi(\bk,t_1)\Big|\hat\Lambda(0,t_1-t_2)\Big|\Psi(\bk',t_2)\otimes\Psi^{\dagger}(\bk,'t_2)\right)\Psi^{\dagger}_2(\bk_2,0)\Psi^{\vphantom{\dagger}}_2(\bk_2,0)\Big\rangle.\nonumber\\
\end{eqnarray}
The vertex correction shown in Fig.~\ref{fig:bubble} corresponds to pairing the $\Psi_1$'s with one side of the vertex operator, and the $\Psi_2$'s with the other side. The other diagram, that corresponds to the `crossed' pairings, will be discussed later. Taking into account that only the $2-3$ subspace of the vertex operator is non-zero gives
\begin{eqnarray}
i\int \frac{d\varepsilon_1d\varepsilon_2d\omega}{(2\pi)^3} \sum_{i,j=1,2}G_{1i}\left(\varepsilon_1+\omega/2,\bk_1\right)G_{\bar i1}\left(\varepsilon_1-\omega/2,\bk_1\right)G_{j2}\left(\varepsilon_2+\omega/2,\bk_2\right)G_{2\bar j}\left(\varepsilon_2-\omega/2,\bk_2\right)\Lambda_{c\,ij}(0,\omega).
\end{eqnarray}
($\bar 1\equiv 2$, $\bar 2\equiv 1$). Performing the $\varepsilon_{1,2}$ integrals yields the contribution of the vertex correction to $\cC_{\uparrow\downarrow}(\bk_1,\bk_2)$ (and also to $\cC_{\uparrow\uparrow}(\bk_1,\bk_2)$)
\begin{eqnarray} \label{vert_correct}
\tilde\cC_{\uparrow\downarrow}(\bk_1,\bk_2)|=-2\Delta^2\mathrm{Im}\int\frac{d\omega}{2\pi}\frac{1}{\sqrt{\Delta^2+\xi_1^2}\left(4\left(\Delta^2+\xi_1^2\right)-\omega^2\right)}\frac{1}{\sqrt{\Delta^2+\xi_2^2}\left(4\left(\Delta^2+\xi_2^2\right)-\omega^2\right)}
\nonumber\\
 \left(\xi_1\xi_2+\omega^2/4\right)\Lambda_1(0,\omega)+ \left(\xi_1\xi_2-\omega^2/4\right)\Lambda_2(0,\omega).
\end{eqnarray}
\end{widetext}
The result does not depend on the relative angle of $\bk_1$ and $\bk_2$. Now we can check the basic idea: that this contribution cancels the naive one calculated in Eq.~\ref{C_naive} to give a vanishing structure factor at zero wavenumber. First note that after $\xi$ integration only the $\Lambda_1-\Lambda_2$ (phase mode) part of the vertex function contributes. Performing the $\xi$ integrals gives.
\begin{eqnarray} \label{vert_correct_int}
\sum_{\bk_1,\bk_2}\,\tilde\cC_{\uparrow\downarrow}(\bk_1,\bk_2)&=&-2\nu\Delta^2\int _{|\omega|>2\Delta}\frac{d\omega}{2\pi}\frac{1}{|\omega|\sqrt{\omega^2-4\Delta^2}}\nonumber\\&&\qquad\times\cos^{-1}\sqrt{1-\omega^2/4\Delta^2}\nonumber \\
&=&-\frac{\pi\nu\Delta}{4}.
\end{eqnarray}  
With the required factor of 4 for the spin sum (\ref{vert_correct_int}) exactly cancels (\ref{struct_fact_naive}). 
%
%
%
Turning now to the contribution of finite momentum Cooper pairs, we will see that the diagram in Fig.~\ref{fig:nq}a is the leading contribution to $\cC_{\uparrow\downarrow}(\bk_1\to -\bk_2)$. This diagram simply corresponds to the `crossed' pairing in (\ref{first_order}). We can ignore vertex corrections, as we know that they give rise only to a background independent of the angle between $\bk_1$ and $\bk_2$, so in particular don't contribute to any singular contribution. 
The resulting expression is in general rather lengthy. Note that we now require the vertex function at finite $\bq=\bk_1+\bk_2$. At low $\bq$, however, the final $\Omega$ integral is dominated by the singular residue of the phase mode $\Lambda_1(\bq,\omega)-\Lambda_2(\bq,\omega)$
\[\mathrm{Res}\left[\Lambda_1(\bq,\omega)-\Lambda_2(\bq,\omega\right]_{\omega=\pm c_s q}=\pm\frac{2mc_s\Delta^2}{n_p q}\]
where $c_s=v_F/\sqrt{3}$ and $n_p=k_F^3/6\pi^2$ is the density of pairs. This singular behavior gives rise to the second term in (\ref{nq_result}). Though we worked in the BCS limit, we believe this form is valid throughout the crossover, with the appropriate sound velocity $c_s$. Note that because we worked at weak coupling, no depletion of the condensate is accounted for. With depletion, both terms in (\ref{nq_result}) should be scaled by the condensate fraction.

The author would like to thank Victor Gurarie, Leo Radzihovsky and Dan Sheehy for discussions, and for their hospitality during a visit to University of Colorado, where this work was completed.


\begin{thebibliography}{15}
\expandafter\ifx\csname natexlab\endcsname\relax\def\natexlab#1{#1}\fi
\expandafter\ifx\csname bibnamefont\endcsname\relax
  \def\bibnamefont#1{#1}\fi
\expandafter\ifx\csname bibfnamefont\endcsname\relax
  \def\bibfnamefont#1{#1}\fi
\expandafter\ifx\csname citenamefont\endcsname\relax
  \def\citenamefont#1{#1}\fi
\expandafter\ifx\csname url\endcsname\relax
  \def\url#1{\texttt{#1}}\fi
\expandafter\ifx\csname urlprefix\endcsname\relax\def\urlprefix{URL }\fi
\providecommand{\bibinfo}[2]{#2}
\providecommand{\eprint}[2][]{\url{#2}}

\bibitem[{\citenamefont{Regal et~al.}(2004)\citenamefont{Regal, Greiner, and
  Jin}}]{regal2004}
\bibinfo{author}{\bibfnamefont{C.~A.} \bibnamefont{Regal}},
  \bibinfo{author}{\bibfnamefont{M.}~\bibnamefont{Greiner}}, \bibnamefont{and}
  \bibinfo{author}{\bibfnamefont{D.~S.} \bibnamefont{Jin}},
  \bibinfo{journal}{Phys. Rev. Lett.} \textbf{\bibinfo{volume}{92}},
  \bibinfo{eid}{040403} (\bibinfo{year}{2004}).

\bibitem[{\citenamefont{Chin et~al.}(2004)\citenamefont{Chin, Bartenstein,
  Altmeyer, Riedl, Jochim, Denschlag, and Grimm}}]{chin2004}
\bibinfo{author}{\bibfnamefont{C.}~\bibnamefont{Chin}},
  \bibinfo{author}{\bibfnamefont{M.}~\bibnamefont{Bartenstein}},
  \bibinfo{author}{\bibfnamefont{A.}~\bibnamefont{Altmeyer}},
  \bibinfo{author}{\bibfnamefont{S.}~\bibnamefont{Riedl}},
  \bibinfo{author}{\bibfnamefont{S.}~\bibnamefont{Jochim}},
  \bibinfo{author}{\bibfnamefont{J.~H.} \bibnamefont{Denschlag}},
  \bibnamefont{and} \bibinfo{author}{\bibfnamefont{R.}~\bibnamefont{Grimm}},
  \bibinfo{journal}{Science} \textbf{\bibinfo{volume}{305}},
  \bibinfo{pages}{1128} (\bibinfo{year}{2004}).

\bibitem[{\citenamefont{Zwierlein et~al.}(2005)\citenamefont{Zwierlein,
  Abo-Shaeer, Schirotzek, Schunck, and Ketterle}}]{zwierlein2005}
\bibinfo{author}{\bibfnamefont{M.~W.} \bibnamefont{Zwierlein}},
  \bibinfo{author}{\bibfnamefont{J.~R.} \bibnamefont{Abo-Shaeer}},
  \bibinfo{author}{\bibfnamefont{A.}~\bibnamefont{Schirotzek}},
  \bibinfo{author}{\bibfnamefont{C.~H.} \bibnamefont{Schunck}},
  \bibnamefont{and} \bibinfo{author}{\bibfnamefont{W.}~\bibnamefont{Ketterle}},
  \bibinfo{journal}{Nature} \textbf{\bibinfo{volume}{45}},
  \bibinfo{pages}{1047} (\bibinfo{year}{2005}).

\bibitem[{\citenamefont{Regal et~al.}(2005)\citenamefont{Regal, Greiner,
  Giorgini, Holland, and Jin}}]{regal2005}
\bibinfo{author}{\bibfnamefont{C.~A.} \bibnamefont{Regal}},
  \bibinfo{author}{\bibfnamefont{M.}~\bibnamefont{Greiner}},
  \bibinfo{author}{\bibfnamefont{S.}~\bibnamefont{Giorgini}},
  \bibinfo{author}{\bibfnamefont{M.}~\bibnamefont{Holland}}, \bibnamefont{and}
  \bibinfo{author}{\bibfnamefont{D.~S.} \bibnamefont{Jin}},
  \bibinfo{journal}{cond-mat/0507316}  (\bibinfo{year}{2005}).

\bibitem[{\citenamefont{Altman et~al.}(2004)\citenamefont{Altman, Demler, and
  Lukin}}]{altman2004}
\bibinfo{author}{\bibfnamefont{E.}~\bibnamefont{Altman}},
  \bibinfo{author}{\bibfnamefont{E.}~\bibnamefont{Demler}}, \bibnamefont{and}
  \bibinfo{author}{\bibfnamefont{M.~D.} \bibnamefont{Lukin}},
  \bibinfo{journal}{Phys. Rev. A} \textbf{\bibinfo{volume}{70}},
  \bibinfo{eid}{013603} (\bibinfo{year}{2004}).

\bibitem[{\citenamefont{Greiner et~al.}(2005)\citenamefont{Greiner, Regal,
  Stewart, and Jin}}]{greiner2005}
\bibinfo{author}{\bibfnamefont{M.}~\bibnamefont{Greiner}},
  \bibinfo{author}{\bibfnamefont{C.~A.} \bibnamefont{Regal}},
  \bibinfo{author}{\bibfnamefont{J.~T.} \bibnamefont{Stewart}},
  \bibnamefont{and} \bibinfo{author}{\bibfnamefont{D.~S.} \bibnamefont{Jin}},
  \bibinfo{journal}{Phys. Rev. Lett.} \textbf{\bibinfo{volume}{94}},
  \bibinfo{eid}{110401} (\bibinfo{year}{2005}).

\bibitem[{\citenamefont{F\"olling et~al.}(2005)\citenamefont{F\"olling,
  Gerbier, Widera, Mandel, Gericke, and Bloch}}]{folling2005}
\bibinfo{author}{\bibfnamefont{S.}~\bibnamefont{F\"olling}},
  \bibinfo{author}{\bibfnamefont{F.}~\bibnamefont{Gerbier}},
  \bibinfo{author}{\bibfnamefont{A.}~\bibnamefont{Widera}},
  \bibinfo{author}{\bibfnamefont{O.}~\bibnamefont{Mandel}},
  \bibinfo{author}{\bibfnamefont{T.}~\bibnamefont{Gericke}}, \bibnamefont{and}
  \bibinfo{author}{\bibfnamefont{I.}~\bibnamefont{Bloch}},
  \bibinfo{journal}{Nature} \textbf{\bibinfo{volume}{45}},
  \bibinfo{pages}{1047} (\bibinfo{year}{2005}).

\bibitem[{\citenamefont{Price}(1954)}]{price1954}
\bibinfo{author}{\bibfnamefont{P.~J.} \bibnamefont{Price}},
  \bibinfo{journal}{Phys. Rev.} \textbf{\bibinfo{volume}{94}},
  \bibinfo{pages}{257} (\bibinfo{year}{1954}).

\bibitem[{\citenamefont{Buchler et~al.}(2004)\citenamefont{Buchler, Zoller, and
  Zwerger}}]{buchler2004}
\bibinfo{author}{\bibfnamefont{H.~P.} \bibnamefont{Buchler}},
  \bibinfo{author}{\bibfnamefont{P.}~\bibnamefont{Zoller}}, \bibnamefont{and}
  \bibinfo{author}{\bibfnamefont{W.}~\bibnamefont{Zwerger}},
  \bibinfo{journal}{Phys. Rev. Lett.} \textbf{\bibinfo{volume}{93}},
  \bibinfo{eid}{080401} (\bibinfo{year}{2004}).

\bibitem[{\citenamefont{Anderson}(1958)}]{anderson1958}
\bibinfo{author}{\bibfnamefont{P.~W.} \bibnamefont{Anderson}},
  \bibinfo{journal}{Phys. Rev.} \textbf{\bibinfo{volume}{110}},
  \bibinfo{pages}{827} (\bibinfo{year}{1958}).

\bibitem[{\citenamefont{Schreiffer}(1964)}]{schreiffer_book}
\bibinfo{author}{\bibfnamefont{J.~R.} \bibnamefont{Schreiffer}},
  \emph{\bibinfo{title}{Theory of Superconductivity}}
  (\bibinfo{publisher}{Perseus Books, Reading, MA}, \bibinfo{year}{1964}).

\bibitem[{\citenamefont{Engelbrecht et~al.}(1997)\citenamefont{Engelbrecht,
  Randeria, and {S\'a de Melo}}}]{engelbrecht1997}
\bibinfo{author}{\bibfnamefont{J.~R.} \bibnamefont{Engelbrecht}},
  \bibinfo{author}{\bibfnamefont{M.}~\bibnamefont{Randeria}}, \bibnamefont{and}
  \bibinfo{author}{\bibfnamefont{C.~A.~R.} \bibnamefont{{S\'a de Melo}}},
  \bibinfo{journal}{Phys. Rev. B} \textbf{\bibinfo{volume}{55}},
  \bibinfo{pages}{15153} (\bibinfo{year}{1997}).

\bibitem[{\citenamefont{Gavoret and Nozieres}(1964)}]{gavoret1964}
\bibinfo{author}{\bibfnamefont{J.}~\bibnamefont{Gavoret}} \bibnamefont{and}
  \bibinfo{author}{\bibfnamefont{P.}~\bibnamefont{Nozieres}},
  \bibinfo{journal}{Ann. Phys. (NY)} \textbf{\bibinfo{volume}{28}},
  \bibinfo{pages}{349} (\bibinfo{year}{1964}).

\bibitem[{\citenamefont{Chester and Reatto}(1966)}]{chester1966}
\bibinfo{author}{\bibfnamefont{G.~V.} \bibnamefont{Chester}} \bibnamefont{and}
  \bibinfo{author}{\bibfnamefont{L.}~\bibnamefont{Reatto}},
  \bibinfo{journal}{Phys. Lett} \textbf{\bibinfo{volume}{22}},
  \bibinfo{pages}{276} (\bibinfo{year}{1966}).

\bibitem[{\citenamefont{Pitaevskii and Stringari}(1991)}]{pitaevskii1991}
\bibinfo{author}{\bibfnamefont{L.}~\bibnamefont{Pitaevskii}} \bibnamefont{and}
  \bibinfo{author}{\bibfnamefont{S.}~\bibnamefont{Stringari}},
  \bibinfo{journal}{J. Low Temp. Phys.} \textbf{\bibinfo{volume}{85}},
  \bibinfo{pages}{377} (\bibinfo{year}{1991}).

\end{thebibliography}
 \end{document}